\documentclass[fleqn,usenatbib]{mnras}

\usepackage{newtxtext,newtxmath}

\usepackage[T1]{fontenc}
\usepackage{ae,aecompl}

\usepackage{graphicx}	
\usepackage{amssymb}	
\usepackage{times}
\newcommand{\msun}{$ {\rm M}_{\odot}/h$}
\newcommand{\mc}{$ M_{200c}$}
\newcommand{\rc}{$ R_{200c}$}
\newcommand{\zinfall}{$z_{\rm infall}$}

\title[Chemical pre-processing in IllustrisTNG]{Chemical pre-processing of cluster galaxies over the past 10 billion years in the IllustrisTNG simulations}

\author[Gupta et al.]{
Anshu Gupta$^{1,2}$\thanks{E-mail: anshu.gupta@anu.edu.au (AG)},
Tiantian Yuan$^{2,3}$\thanks{ASTRO 3D Fellow},
Paul Torrey$^{4}$\thanks{Hubble Fellow},
Mark Vogelsberger$^{4}$\thanks{Alfred P. Sloan Fellow},
\newauthor{
Davide Martizzi$^5$,
Kim-Vy H. Tran$^{2,6,7,8}$,
Lisa J. Kewley$^{1,2}$,
Federico Marinacci$^4$,
}
\newauthor{
Dylan Nelson$^9$,
Annalisa Pillepich$^{10,11}$,
Lars Hernquist$^{10}$,
Shy Genel$^{12,13}$,
}
\newauthor{
Volker Springel$^{14,15,9}$
}\\
$^1$Research School of Astronomy and Astrophysics, Australian National University, Canberra, ACT 2611, Australia \\
$^2$ARC Centre of Excellence for All Sky Astrophysics in 3 Dimensions (ASTRO 3D), Australia\\
$^3$Centre for Astrophysics and Supercomputing, Swinburne University of Technology, Hawthorn, Victoria 3122, Australia \\
$^4$MIT Kavli Institute for Astrophysics \& Space Research, Cambridge, MA, 02139, USA\\
$^5$Department of Astronomy, University of California, Berkeley, CA 94720-3411, USA\\
$^6$School of Physics, University of New South Wales, Kensington,Australia\\
$^7$Australian Astronomical Observatory\\
$^8$George P. and Cynthia W. Mitchell Institute for Fundamental Physics and Astronomy, Department of Physics \& Astronomy, \\ Texas A\&M University, College Station, TX 77843\\
$^9$Max-Planck-Institut f\"{u}r Astrophysik, Karl-Schwarzschild-Str. 1, 85741 Garching, Germany\\
$^{10}$Harvard-Smithsonian Center for Astrophysics, 60 Garden Street, Cambridge, MA 02138\\
$^{11}$Max-Planck-Institut f\"{u}r Astronomie, K\"{o}nigstuhl 17, 69117 Heidelberg, Germany\\
$^{12}$Center for Computational Astrophysics, Flatiron Institute, 162 Fifth Avenue, New York, NY 10010, USA\\
$^{13}$Department of Astronomy, Columbia University, 550West 120th Street, New York, NY 10027, USA\\
$^{14}$Heidelberger Institut f\"{u}r Theoretische Studien, Schloss-Wolfsbrunnenweg 35, 69118 Heidelberg, Germany\\
$^{15}$Zentrum f\"{u}r Astronomie der Universit\"{a}t Heidelberg, Astronomisches Recheninstitut, M\"{o}nchhofstr. 12-14, 69120, Heidelberg, Germany
}

\date{Accepted XXX. Received YYY; in original form ZZZ}

\pubyear{2017}

\begin{document}

\label{firstpage}
\pagerange{\pageref{firstpage}--\pageref{lastpage}}
\maketitle

\begin{abstract}

We use the IllustrisTNG simulations to investigate the evolution of the mass-metallicity relation (MZR) for star-forming cluster galaxies as a function of the formation history of their cluster host. The simulations predict an enhancement in the gas-phase metallicities of star-forming cluster galaxies ($10^9<  M_*<10^{10}\,$\msun) at $z\leq1.0$ in comparisons to field galaxies. This is qualitatively consistent with observations.  We find that the metallicity enhancement of cluster galaxies appears prior to their infall into the central cluster potential, indicating for the first time a systematic ``chemical pre-processing'' signature for {\it infalling} cluster galaxies. Namely, galaxies which will fall into a cluster by $z=0$ show a $\sim0.05$\,dex enhancement in the MZR compared to field galaxies at $z\leq0.5$. Based on the inflow rate of gas into cluster galaxies and its metallicity, we identify that the accretion of pre-enriched gas is the key driver of the chemical evolution of such galaxies, particularly in the stellar mass range ($10^9<  M_*<10^{10}\,$\msun). We see signatures of an environmental dependence of the ambient/inflowing gas metallicity which extends well outside the nominal virial radius of clusters. Our results motivate future observations looking for pre-enrichment signatures in dense environments.

\end{abstract}




\section{Introduction}
The chemical abundance of galaxies encodes the cumulative history of baryonic processes such as star formation and gas inflow/outflow. Observations clearly show that the cluster environment causes morphological and color transformations of galaxies \citep[e.g.,][]{Dressler1980}. However, the impact of cluster-scale overdensities on the chemical evolution of galaxies remains controversial.   Observations at $z\sim0$ find a minimal difference (at most 0.05 dex) in the mass-metallicity relation (MZR) of  cluster and field  star-forming (SF) galaxies \citep{Ellison2009, Pasquali2012, Pilyugin2016a}.  On the other hand, significant dependencies of the gas-phase metallicity on the cluster-centric distance and the cluster dynamic state are  observed \citep{Petropoulou2012, Gupta2016}.  Observations of the MZR at $z>1.0$ are  reported for only a handful of clusters with conflicting results \citep{Tran2015, Valentino2015, Kacprzak2015}.

In a hierarchical $\Lambda$CDM Universe, massive clusters form through the merger of smaller galaxy groups.  Physical properties of  group galaxies may be modified  prior to accretion onto the main cluster,  a process referred to as  ``pre-processing'' \citep[e.g.,][]{Zabludoff1998, Hou2014}.  The most recent observational evidence of pre-processing includes H\,I-deficiency and quenching of SF  in galaxy groups \citep{Brown2017, Bianconi2017}, though the significance of group pre-processing is debatable \citep[e.g.,][]{Berrier2009}.  \cite{Darvish2015} find $\sim 0.10$\,dex higher metallicity for galaxies in a filamentary structure at $z\sim0.5$ than the counterpart field galaxies, suggesting a pre-processing role of galaxy filaments on the metallicity.  Whether, when and how the chemical abundance of galaxies is altered in the cluster formation history remains an open question. 
 
Current observations are plagued by poorly understood systematic errors such as projection effects, interlopers, selection biases and metallicity diagnostics.  These errors can easily overwhelm the chemical enrichment signal as a function of environment, particularly at high redshifts. In addition, the diverse definition of environment makes comparisons across different studies ambiguous \citep{Ellison2009, Peng2014}. Cosmological simulations with a full 3-dimensional formation history of cluster galaxies are a powerful tool  to overcome  observational biases and understand the physics of the environmental dependence of the chemical evolution \citep{Dave2011, Genel2016, Bahe2017}. 

In this Letter, we explore the MZR evolution as a function of the formation history of galaxy groups and clusters in IllustrisTNG, a large-scale cosmological simulation. Using a novel method of separating progenitor cluster galaxies into accreted and  infalling categories, we report for the first time a clear systematic signature of ``chemical pre-processing'' in cluster galaxies over the past 10 Gyrs.  We show the effectiveness of pre-enriched gas inflows in driving the chemical evolution of cluster galaxies  out to twice the virial radius. 

\begin{figure}
\centering
\tiny
\includegraphics[scale=0.38, trim=3.5cm 3.5cm 2.5cm 2.5cm,clip=true]{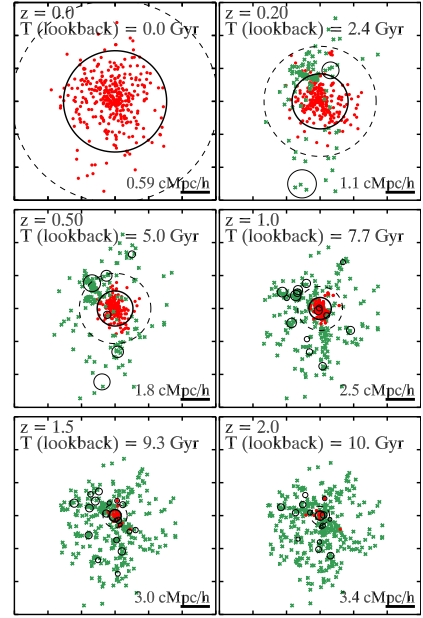}
\caption{The assembly history of the most massive galaxy cluster ($10^{14.4}\,$\msun) in the TNG100  simulation from the projected spatial distribution of cluster members. The red circles and green asterisks represent accreted (\zinfall $\,>\,z$) and infalling (\zinfall $\,<\,z$) cluster members respectively. The solid black circles corresponds to the \rc\ of the dark matter halos associated with the progenitor cluster galaxies. The dashed-black circle represents the cluster boundary ($2\times$\,\rc) at each snapshot.  All distances are represented in projected comoving scale. Throughout this Letter we present the evolution of physical properties of galaxies in six redshift slices $z = [0.0,0.2,0.5,1.0,1.5,2.0]$.}
\label{fig:subhalo_dis}
\end{figure}

\section{Methods}\label{sec:sample}
Our results are based on the IllustrisTNG simulations~\citep[][]{Pillepich2017, Nelson2017, Springel2017, Marinacci2017, Naiman2017}. The galaxy formation model is an updated and extended version of the Illustris model~\citep[][]{Vogelsberger2013, Torrey2014}. Details of the model modifications are described in \cite{Weinberger2016} and \citet{Pillepich2017a}.  We use the TNG100 simulation that has a volume of  $\sim(100\,{\rm Mpc})^3$ and $ m_b = 9.4 \times 10^5$\,\msun, where $m_b$ is the baryonic  mass per particle.  


\begin{figure}
\centering
\tiny
\includegraphics[scale=0.40, trim=0.6cm 1.7cm 2.5cm 2.5cm,clip=true]{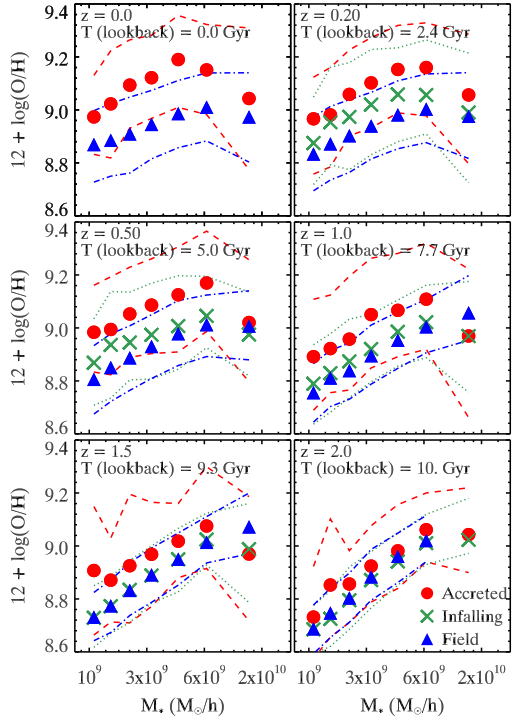}
\caption{Evolution of the mass-metallicity relation for the accreted cluster (red circles), infalling cluster (green asterisks) and field galaxies (blue triangle). At any given redshift, we are plotting the mean SFR-weighted gas-phase metallicity of galaxies with SFR$>0$. The coloured dashed and dotted lines represents the 16th and 84th percentiles. The accreted cluster members show nearly $0.10-0.18$\,dex metal enhancement with respect to field galaxies at each stellar mass bin at $z\le1.5$. The infalling cluster members show $0.05-0.08$\,dex metal enhancement with respect to field galaxies at $z\le0.5$ suggesting chemical pre-processing of infalling galaxies.}
\label{fig:mz_evo}
\end{figure}

 From the TNG100 simulation, we select all halos with \mc\,$\ge\,10^{13.0}\,$\msun\, where \mc\ is total mass enclosed within \rc\ (the radius within which the mean density is 200 times the critical density of the Universe at the halo redshift). This leads to a selection of 127 galaxy clusters and groups at $z=0$. The mean cluster halo mass of our sample is $10^{13.4}\,$\msun. We define the boundary of a  cluster as twice the \rc\ of the cluster halo, which is  similar to the splashback radius \citep{Diemer2017}. Out of all Friends-of-Friends (FOF) subhalos associated with the cluster at $z=0$,   all galaxies whether centrals or satellites  residing within twice the \rc\  make up our cluster member galaxy sample. Note that we exclude  flybys and satellite galaxies in the process of splashing back at $z=0$ from our analysis. 

We restrict this analysis to a stellar mass range of $10^9-10^{10.5}$\,\msun. The lower stellar mass cut is to minimize numerical uncertainties such that there are at least 1000 stellar particles per galaxy (a single star particle in TNG100 has a mass of about $10^6$\,\msun).   We impose the upper mass cut of $<10^{10.5}\,$\msun\ to reduce contaminations from active galactic nuclei (AGN) activities \citep{Pillepich2017}.  The removal of high-mass star-forming ($M_*>10^{10.5}\,$\msun) galaxies  does not change our results significantly because  environmental effects are more efficient for low mass galaxies. We select 3567 cluster member galaxies with stellar mass $10^9-10^{10.5}$\,\msun, without any star formation rates (SFR) cut at $z=0$. Most cluster galaxies at $z=0$ are not forming stars.  

Our field galaxy sample consists of 8900 galaxies at $z=0$ that reside in host halos of  mass \mc\,$<\,10^{12.0}\,$\msun, have SFR\,$>0$, and  $10^9<M_*<10^{10.5}$\,\msun. We focus on the SFR-weighted gas-phase metallicity represented in terms of fractional oxygen abundance ($12+\log({\rm O/H})$) to facilitate a direct comparison with the nebular metallicity observations. Throughout this Letter the stellar mass and SFR  are measured within twice the stellar half mass radius.  Our SFR-weighted gas metallicity is independent of aperture and has been found to provide a good approximation of the global metallicity from observations \citep{Dave2011, Bahe2017,  Torrey2017}. 
 
\begin{figure}
\centering
\tiny
\includegraphics[scale=0.40, trim=0.6cm 1.7cm 2.5cm 2.5cm,clip=true]{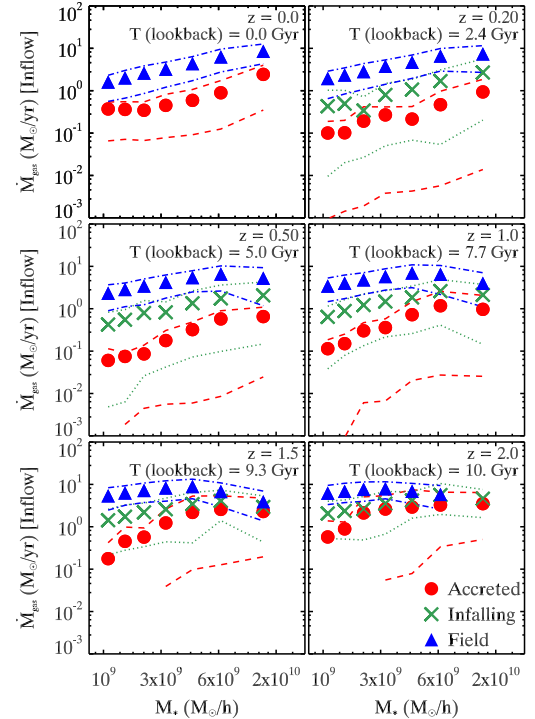}
\caption{Gas mass inflow rate versus stellar mass for the accreted cluster (red circles), infalling cluster (green asterisks) and field (blue triangles) galaxies.  Only star-forming galaxies (SFR$>0$) are included in the analysis at any given redshift. The data points represent the  mean gas mass inflow rate, and the lines indicate the 16th and 84th percentiles for the respective sample.  At each redshift snapshot,  cluster member galaxies have consistently lower gas inflow rates  compared to field galaxies.  }
\label{fig:tracer_inflow}
\end{figure}

We track {\em all} $z=0$ cluster and field galaxies back in time, but galaxies with stellar mass between $10^9-10^{10.5}\,$\msun\ and SFR$\,>0$ at a given redshift, will only be shown in Figure \ref{fig:mz_evo}, \ref{fig:tracer_inflow} and \ref{fig:tracer_inflow_met}. We use merger tree catalog based on the technique of \citet{Rodriguez-gomez2015}  to track progenitors of the cluster and field galaxy samples at $z=0$. At each redshift slice, we use the host halo of the progenitor central cluster galaxy to identify the cluster center and the cluster boundary based on its \rc.  We estimate \zinfall\  for each progenitor of the present-day cluster member galaxy as the earliest time (highest redshift) at which the galaxy crosses the cluster boundary ($R < 2\times\,$\rc\ of the central cluster halo) for the first time.  Using \zinfall, we separate progenitor cluster galaxies into two subgroups at each redshift slice:  accreted (\zinfall $\,>\,z$) and infalling (\zinfall $\,<\,z$) galaxies. 

This technique can unambiguously separate the infalling galaxies from the accreted cluster galaxies that are on their second passage into the cluster potential. At each redshift, the progenitors of the field galaxy sample form our comparison sample.   Figure \ref{fig:subhalo_dis} shows the assembly history of the most massive cluster in  TNG100 (\mc\,$ = 10^{14.4}$\,\msun) through the projected spatial distribution of progenitors at a few redshift snapshots.  At high redshift ($z\sim1.5$),  progenitor cluster galaxies have a comoving spatial extent of  $\sim 10-20\, {\rm Mpc}/h$, consistent with the recent work by \cite{Chiang2017}.  


\section{Results}
\subsection{Cosmic evolution of the mass-metallicity relation with environment}\label{sec:mzr_evo}
Figure \ref{fig:mz_evo} shows  the stellar mass versus SFR-weighted gas-phase metallicity for our three galaxy samples (accreted cluster, infalling cluster, and field) at six different redshifts.    We restrict our gas metallicity analysis only to  galaxies with SFR\,$ >0$ at any given redshift.   We bin the data by stellar mass such that each stellar mass bin has nearly the same number of galaxies at $z=0$.  At any given redshift, we use the same stellar mass bins defined at $z=0$ and only plot bins that have at least 10 galaxies. The MZR of accreted and infalling  galaxies is consistent with the MZR of the field galaxy sample at $z=2.0$. Signs of the environmental dependence of metallicity emerge around $z=1.5$, in particular the low mass accreted galaxies ($<10^{9.5}\,$\msun)   are $\sim 0.05$\,dex  more metal-rich compared to counterpart field galaxies. At $z\leq1.0$, the gas-phase metallicity of accreted cluster galaxies show a consistent metallicity enhancement of $0.15-0.20\,$dex with respect to field galaxies.  

The MZR of infalling galaxies  lies persistently in between accreted and field galaxies for $z \sim 0-2$, signalling the action of chemical pre-processing in infalling galaxies before they are fully accreted onto their respective clusters. The low mass infalling galaxies ($<10^{9.5}\,$\msun) at $z=0.5$ show an offset of $\sim 0.05\,$dex from the counterpart field galaxies. The metal enhancement permeates to even higher mass infalling galaxies ($<10^{10}\,$\msun) at $z\sim0.2$. The infalling galaxies have not crossed the cluster boundary ($2\times$\,\rc), hence the metallicity offset between infalling cluster and field galaxies appears prior to infall.  Our MZR analysis shows for the first time a systematic  signature of  chemical pre-processing in infalling galaxies. In this Letter, we use the term pre-processing more broadly to describe the modification in the properties of galaxies prior to infall onto the main cluster irrespective of their prior group membership.

The MRZ evolution of cluster and field shows a qualitative agreement with observations \citep{Ellison2009, Pilyugin2016a}, though a detailed quantitative comparison with observations is not feasible because of discrepancies in emission line diagnostics \citep{Kewley2008},{\bf  the choice of nucleosynthesis yields in the simulations,} diverse definition of environments,  and selection effects.  Most observations find a maximum of 0.05\,dex metallicity enhancement for cluster galaxies \citep{Pilyugin2016a}, significantly lower than the one predicted by simulations. Contamination by interlopers or infalling galaxies due to projection effects and the lack of dynamic time estimates can easily wash out the metallicity enhancement in observations. Also, a higher SFR cut in observations specifically removes cluster galaxies undergoing change due to the environment. We test different SFR cuts and  find a maximum SFR cut of $0.1\,$\msun$/{\rm yr}$ is required to observe the environment driven changes in the metallicity, any higher SFR cut will dilute the observed metallicity enhancement. 

The simulation  shows a 1-sigma scatter of $\sim0.20\,$dex in the MZR.  The large scatter underlines the cluster-to-cluster variation and shows the challenge of detecting the metallicity enhancement with relatively small cluster samples in observations. Meanwhile, we notice that the scatter of the simulated MZR at $z\sim0$ seems to be large in comparison with observations \citep[e.g.,][]{Pilyugin2016a}.  We plan to investigate the origin of the scatter in more detail in a future work. Simulations predict no significant evolution in the average metallicity of the highest mass bin ($10^{10}-10^{10.5}\,$\msun) between $z=2.0\ {\rm to}\ z=0.0$ for any galaxy sample. The average metallicity in the  highest mass bin is biased towards  galaxies with either extremely low SFRs and/or significant AGN contribution. 

\subsection{Properties of gas accretion in high-density environment}\label{sec:gas_inflow}
To understand the origin of the environmental dependence of the MZR and the chemical pre-processing,  we investigate the gas accretion history of our three samples using Monte Carlo tracer particles \citep{Genel2013}. At any given redshift, we measure the tracer flux associated with baryons entering the galaxy for the first time over a time interval of 1 Gyr. We then multiply the tracer count with an effective tracer mass to estimate a ``raw gas mass inflow rate'' \citep{Nelson2015c}.    The median number of tracer particles associated with the gas mass inflow rate is about $800-15000$. 


Figure \ref{fig:tracer_inflow} shows the average gas mass inflow rate for the three samples (accreted, infalling and field) in six redshift slices.  We find that the gas mass inflow rate is consistently lower by $0.4-1.0$\,dex  for accreted and infalling cluster galaxies compared to field galaxies across all redshift slices. The difference in the average gas mass inflow rate  between cluster (accreted and infalling) and field galaxies increases slightly with cosmic time, i.e., increases from $\sim 0.4$\,dex at $z=2.0$ to  $\sim1.0$\,dex at $z<0.5$. Almost 50\% of  accreted cluster galaxies with non-zero gas mass inflow rate and SFRs have been accreted onto the main cluster in the past 2 Gyrs. In TNG100, cluster galaxies have on average 0.5\,dex lower SFRs than field galaxies at all epochs. \cite{Torrey2017} show  that the MZR of field galaxies evolves through either the accretion dominant or enrichment dominant phase.   
\begin{figure}
\centering
\tiny
\includegraphics[scale=0.40, trim=1.1cm 1.5cm 1.0cm 1.5cm,clip=true]{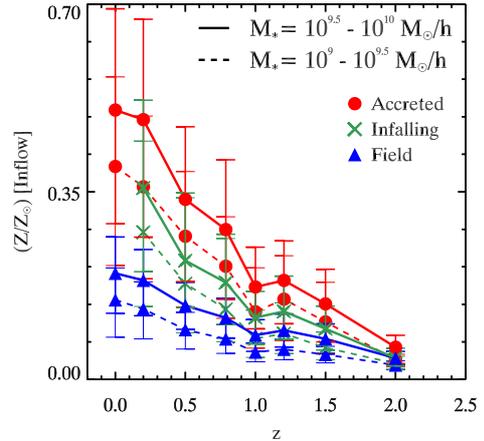}
\caption{Cosmic evolution of the mean metallicity of the inflowing gas for cluster (red), infalling (green) and field (blue) galaxies. The solid and dashed lines represent galaxies in stellar mass bins labeled in the figure.   The inflowing gas for both accreted and infalling cluster galaxies is $2-3$ times more metal rich at $z=0$ than field galaxies. Note that we are plotting the total metal content not the fractional oxygen abundance. The error bars represent 16th and 84th percentiles. The solar metallicity is $Z_{\odot} = 0.0134$ \citep{Asplund2009}. }
\label{fig:tracer_inflow_met}
\end{figure}

Figure \ref{fig:tracer_inflow_met} shows the redshift evolution of the mean metallicity of the inflowing gas. The average metallicity of the baryons associated with the raw gas mass inflow rate is defined as the mean metallicity of inflowing gas. The mass abundance of all metals used in Figure \ref{fig:tracer_inflow_met} relates to the oxygen number abundance by a simple multiplicative factor ( $\sim 0.03$). The average metallicity of the inflowing gas increases with cosmic time,  irrespective of the galaxy sample or the stellar mass bin. However, galaxies in the cluster environment (accreted or infalling) receive more metal-rich inflowing gas than galaxies in the field, and relatively more so at lower redshifts (Figures \ref{fig:tracer_inflow_met}).   Even at $z = 1.5$, accreted cluster galaxies are receiving almost 1.3 times more metal-rich gas than field galaxies; at $z \sim 0$ the difference in metal content can be as large as a factor of 3. Yet, the inflowing gas mass rate is $\sim1.0$ dex lower for cluster galaxies than field galaxies.   We also find  a  secondary positive correlation in the metallicity of inflowing gas and the stellar mass,  consistent with previous suggestions \citep{Fraternali2016, Kacprzak2016}.   

The average metallicity of the inflowing gas for infalling galaxies continues to lie in-between accreted cluster and field galaxies at all epochs (Figure \ref{fig:tracer_inflow_met}).  In fact, at $z<0.5$,  infalling galaxies are accreting $1.5-2$ times more metal-rich gas compared to field galaxies.  The infalling galaxies have never crossed the boundary of the central cluster potential ($2\times$\,\rc) and are receiving metal-rich gas compared to field galaxies, suggesting that enriched gas inflows can be significant at  distances larger than $2\times$\,\rc.  Figure \ref{fig:tracer_inflow} and  Figure \ref{fig:tracer_inflow_met}  suggest that the enriched gas inflows contribute significantly to the chemical enrichment of galaxies in overdense environments.

\section{Discussion and Conclusion}\label{sec:conclusions}

Using the IllustrisTNG cosmological simulation, we demonstrate the first systematic signature of ``chemical pre-processing'' of infalling cluster galaxies, namely galaxies which will be accreted by $z=0$ into a massive host. Before infall, these galaxies exhibit a $\sim0.05$\,dex enhancement in the MZR in comparison to field galaxies, a difference  observable at low redshifts ($z < 0.5$, Figure \ref{fig:mz_evo}). The simulation also predicts a systematic pre-enrichment of the gas which is available for inflow into cluster galaxies (Figure \ref{fig:tracer_inflow_met}). At $z < 1.0$, cluster galaxies (both already accreted and infalling) accrete gas which is $2-3$ times more metal-rich compared to field galaxies. 


We suggest that infalling galaxies exhibiting pre-processing signatures may or may not belong to  smaller, infalling groups, a distinction that will be checked in a follow-up paper.  The underlying pre-processing physical mechanisms do not act necessarily on the cluster galaxies themselves, but rather on the gas around them which is available for accretion. Gas removal via gravitational or hydrodynamic interactions in infalling groups has instead previously been  suspected as the cause of  pre-processing of infalling galaxies \citep{Cortese2006, Scott2015}.  Our definition of cluster boundary to separate infalling and accreted cluster galaxies is similar to the splashback radius  \citep{Diemer2017}. The signature of chemical pre-processing can be used as an observational identification for the splashback radius.


The metallicity enhancement of cluster galaxies on the MZR is qualitatively consistent with observations and  recent investigations of cosmological simulations \citep{Dave2011, Genel2016,  Bahe2017}.  \cite{Genel2016} attribute the metallicity enhancement to either the reduced inflow of metal-poor gas or SFR concentration towards the inner, more metal-rich parts for galaxies with stellar mass $\sim10^{10.3}$\,\msun. Observations show that environmental processes are mostly effective for galaxies with $ {\rm M}_*<10^{10}\,$\msun\  \citep{Peng2010}. \cite{Gupta2017} shows that star formation suppression in the galactic outskirts due to ram pressure stripping (RPS) can not directly produce a significant enhancement in the SFR-weighted metallicity. However, a relative increase in the metallicity of cluster galaxies compared to field galaxies due to suppression of pristine gas inflow by RPS can  not be ruled out. We also consistently find lower  gas mass inflow rates for both infalling and accreted cluster galaxies compared to field galaxies in the simulation.

The pre-enrichment of the inflowing gas in clusters follows the cosmic evolution of intra-cluster medium (ICM) enrichment \citep{Vogelsberger2017}. \cite{Peng2014} also suggested that the mixing of the inflowing gas with the metal-rich ICM can be responsible for the pre-enrichment of inflowing gas in the cluster environment.  We speculate that the inflow of pre-enriched gas in combination with the suppression of gas inflows  drive the chemical pre-processing of infalling galaxies and the metallicity enhancement of star-forming cluster galaxies. Carefully designed observations are needed to confirm the chemical pre-processing of infalling galaxies and the pre-enrichment signal in overdense environment. 

\section*{Acknowledgements}
PT was supported by NASA through Hubble Fellowship grant HST-HF-51384.001-A. MV acknowledges support from a MIT RSC award, the Alfred P. Sloan Foundation, and by NASA ATP grant NNX17AG29G. L.J.K. gratefully acknowledges support from an Australian Research Council (ARC) Laureate Fellowship (FL150100113). T.Y. acknowledges support from an ASTRO 3D fellowship. K. Tran acknowledges support by the National Science Foundation under Grant Number 1410728. Parts of this research were conducted by the Australian Research Council Centre of Excellence for All Sky Astrophysics in 3 Dimensions (ASTRO 3D), through project number CE170100013. The IllustrisTNG simulations and the ancillary runs were run on  the HazelHen Cray XC40-system (project GCS-ILLU), Stampede supercomputer at TACC/XSEDE (allocation AST140063), at the Hydra and Draco supercomputers at the Max Planck Computing and Data Facility, and on the MIT/Harvard computing facilities supported by FAS and MIT MKI.


\bsp	
\label{lastpage}
\end{document}